\documentstyle[aps,psfig,prbbib]{revtex}

\newcommand{\bvec}[1]{{\mathbf #1}}
\newcommand{\df}[2]{\delta^{(#2)} ( #1 )}

\newcommand{\ind}[2]{\int d^{#1}\bvec{#2}}

\begin{document}

\draft

\widetext

\title{Two-component approach for thermodynamic properties in
diluted magnetic semiconductors}

\author{Malcolm P. Kennett$^1$, Mona Berciu$^2$ and R. N.
Bhatt$^{1,2}$}

\address{$^1$Department of Physics and $^2$Department of
Electrical
Engineering, Princeton University, Princeton, NJ 08544}

\date{\today}

\twocolumn[\hsize\textwidth\columnwidth\hsize\csname@twocolumnfalse\endcsname

\maketitle


\begin{abstract}
We examine the feasibility of a simple description
of
Mn ions in III-V diluted magnetic semiconductors (DMSs)
in terms of two {\em species} (components),
motivated by the expectation that the Mn-hole exchange couplings
are
widely
distributed, especially for low Mn concentrations.  We find,
using
distributions indicated by recent numerical
mean field studies, that
the thermodynamic properties (magnetization, susceptibility, and
specific
heat) cannot be fit by a single coupling as in a homogeneous
model, but
can be
fit well  by a two-component model
with a temperature dependent number of ``strongly'' and
``weakly'' coupled
spins.  This suggests that a two-component description
may be a minimal model for the
interpretation of experimental
measurements of thermodynamic quantities in III-V DMS systems.

\end{abstract}

\pacs{PACS: 75.50.Pp, 75.10.Lp, 75.40.-s, 75.30.Hx}]


\section{Introduction}
\label{Introduction}

Diluted, magnetic semiconductors (DMSs) have been the focus of
intense
study recently due to their potential for use in novel devices
making
use of
both magnetic and conventional
semiconductor properties. \cite{Ohno1,Prinz}  The discovery of
a  magnetic transition
temperature, $T_c$, of 110 K in a sample of Ga$_{1-x}$Mn$_x$As
for $x = 0.053$ has
further spurred efforts to understand the origin and physical
effects that influence the
magnetic properties of these materials. \cite{Matsukura}

It is now reasonably well established that III-V systems such
as Ga$_{1-x}$Mn$_x$As with $x = 0.01 - 0.07$ are
itinerant ferromagnets, in which the Mn ions play the dual
roles of acceptor site and
magnetic ion, and the itinerant carriers are holes, which have
an antiferromagnetic
interaction with the Mn spins.
\cite{DOM,Beschoten,Ohno2,Szczytko,Dietl}
The antiferromagnetic hole-Mn interaction leads to an effective
ferromagnetic interaction
between Mn spins, and gives rise to the ferromagnetic
transition. One experimental fact that may be of importance
is that in these systems 
the number of holes $n_h$ is only a small fraction of the number
of Mn dopants (or Mn spins), $n_{\rm Mn}$, implying that
the system has low carrier density, and is heavily
compensated.

Theoretically, there have been several approaches to trying to
calculate the thermodynamic
and transport properties of these materials -- one approach has
been to look at the effect
of spin waves
\cite{Konig,Kcorrection,Schliemann}
whilst another has been a mean field model including spin-orbit
effects. \cite{DOM}  Both of these
approaches leave out the disorder due to the random positions
of the Mn ions in the sample.
The effect of random positions has been considered in a
numerical mean field theory \cite{Mona,Mona2} for the low density
phase of
Ga$_{1-x}$Mn$_x$As, as well as
in Monte Carlo simulations of the insulating phase of II-VI
DMS, represented by a Heisenberg
model for the Mn and carrier spins. \cite{Xin1,Xin2}  Both of
these
investigations show that the
positional disorder gives rise to a distribution of exchange
couplings between Mn ions and
holes.  Monte Carlo results have also been obtained for a model
in which the hole-Mn coupling is assumed
constant and leads to an effective Mn-Mn interaction where
positional disorder is included. \cite{Boselli}
A recent Monte Carlo study of Ga$_{1-x}$Mn$_x$As using a
kinetic-exchange
model, \cite{SchliemannMC} has appeared whilst this work was
being
written up.

In this paper, we construct the simplest mean field model
that attempts to capture
the effects of disorder in the effective local fields at different
Mn sites.
This disorder arises as a result of the different local
potential for the carrier at different Mn sites. 
At one extreme we consider a simple model of compensation by
antisite defects. In this model, each As antisite defect
is viewed as capturing the holes of two neighboring Mn dopants,
and providing an onsite potential that is significantly different
from Mn sites that are far from such As antisites. This
naturally leads to description of the Mn spins in terms of
two distinct species. This is clearly a caricature, since
for positionally random doping and As antisite defects, there
will be a {\em continuous distribution} of onsite potentials
rather than a bimodal one. However, as we show in the bulk
of our paper, if the distribution is rather wide (as is found
in the mean-field study of Mn impurity bands\cite{Mona,Mona2} ),
such a bimodal distribution
provides a reasonable description of the thermodynamics,
provided we allow the relative weights of the two species
to be temperature dependent according to a simple rule, which
occurs naturally in the analysis.

We
concentrate only on the
carrier-spin exchange part of the Hamiltonian, since a
numerical mean-field
treatment \cite{Mona2} shows that this term captures most of the
condensation energy for the magnetic phase, and the carrier
kinetic energy
changes only weakly  with the onset of ferromagnetism.
As stated earlier, in the regime of interest, the concentration of holes, $n_h$, is
considerably less than the concentration $n_{\rm Mn}$ of Mn ions.
The fluctuations in the local carrier charge density (due to
fluctuations in the occupation number of the impurity
states around Mn sites with different surroundings) are
then represented by a fluctuation of the effective
exchange coupling between the Mn moment and the spin of the carrier.

Our results can be summarized simply as follows :
we find that representing the hole-Mn antiferromagnetic
exchange coupling by a {\em single} parameter is insufficient to
capture
the
thermodynamic behaviour, such as the temperature dependence of
the
magnetization, susceptibility and specific heat in the
ferromagnetic phase, when disorder is large.
However, by using a
model where there are two species of Mn ions, with
different hole-Mn exchange couplings,
we are able to fit the magnetization and
other thermodynamic parameters that we calculate from a
distribution of couplings
in a much more satisfactory manner.
Furthermore, as argued below, such a two-component model
captures the inadequacies of the homogeneous model
in a manner that is qualitatively correct for temperatures not
too close
to $T_c$.
{\em We emphasize that the label ``component" used throughout
this paper, is synonymous with ``species" or ``type", and has
no relation to the number of components of the spin itself,
which in this work refers to a vector in three
dimensions}.

The idea of using a two-component model is similar in spirit to
models developed to understand the magnetic behaviour of doped
non-magnetic
semiconductors such as phosphorus doped silicon,
motivated by the observation that exchange couplings in such
systems should be
distributed over many orders of magnitude. \cite{Bhatt_Lee,Bhatt}
In Mn doped DMS systems, disorder appears to lead to a
similar situation of
widely distributed couplings. \cite{Mona2}  Whilst not explicitly included
in Refs. \onlinecite{Mona,Mona2} the large
compensation, apparently from As antisite defects, would enhance 
disorder, and hence broaden the distribution of effective couplings.
Therefore a two component approach appears to be
a natural approximation that is qualitatively correct,
and may be quantitatively adequate for many purposes.

\section{Mean Field Model}
\label{mean_field}
\subsection{Hamiltonian}
\label{Ham}

The DMS system we study
consists of magnetic ions (Mn) coupled to charge carriers. In
the case of II-VI semiconductors (e.g. ZnSe), the carrier is
provided by a second dopant (such as P); however, in
III-V semiconductors (e.g. GaAs), Mn, being a divalent
atom substituting on a trivalent (Ga) site, provides a hole
in addition to the spin. In this paper, we use the example of
Ga$_{1-x}$Mn$_x$As for concreteness.

For Mn concentrations of interest (a fraction of a percent
to a few percent), and hole concentrations of about 10\%-20\%
per Mn impurity, the system is near a metal-insulator
transition (MIT). \cite{Ohno2,VanEsch}  This implies that the
hole wavefunctions
are filamentary, with a multifractal structure
on length scales that determine the magnetic behaviour
of the system, which is quite distinct from the homogeneous structure
for plane wave (or Bloch wave) states characteristic of
periodic systems.  Thus, each hole interacts with many
Mn spins, depending on the amplitude of its wavefunction
at various Mn sites, as well as the envelope function
characterizing
the hole. Since each hole interacts with a large number of
spins the net exchange fields
felt by different holes are similar (i.e., the fluctuations
are not that large). However, the hole concentration
is considerably (a factor of 5-10) smaller than the Mn
concentration; therefore, each Mn spin experiences a rather
different exchange field due to the few holes that have
significant amplitude at that site (or nearby sites, via the
tail of
the envelope function). Hence the fluctuations in the local
exchange field at different Mn sites {\it cannot} be ignored,
since the fields are being produced predominantly by
just a few holes. This asymmetry has been documented in
a numerical study \cite{Mona2} and forms the basis of
our simple phenomenological scheme. These fluctuations are
of paramount importance in the insulating phase and at the
MIT. If experience with conventional
doped semiconductors is any guide, they are likely to persist
well into the metallic phase \cite{Mona,Bhatt} and cannot
reasonably be ignored
in any theory of DMS ferromagnetism up to dopant densities
well above (a factor of 3-5) the dopant density for
the MIT.  The above discussion raises questions about the
applicability of the results of studies which
are based on homogeneous mean-field  models, as well as
those based on perturbative treatments of the Mn
spin system (such as RKKY exchange),
given the large ratio of spin to carrier
density.

The diluteness of the carrier system leads us to neglect
hole-hole interactions. Direct Mn-Mn interactions are also
ignored because they are extremely short range due to the
atomic nature of the Mn 3d-orbitals responsible for the Mn
spin. Furthermore, the numerical mean field treatment
\cite{Mona2}
shows that the energy gain due to the onset of ferromagnetic
ordering coming from the exchange term in the Hamiltonian
is much larger than the change in carrier kinetic energy.
Ignoring the kinetic energy is not expected to lead to
any major qualitative differences.
In fact, in our results
it does not lead to even {\em quantitative}
differences in the magnetization for temperatures below about
$0.6 \, T_c$ (as shown in Section \ref{numerical}), while there
are quantitative differences nearer the ordering
temperature, as the exchange energy and
kinetic energy variations with $T$ become comparable in magnitude.
Consequently, in keeping with the philosophy of finding
a minimal model description, we consider only the exchange
term of the Hamiltonian in this paper. This Hamiltonian
takes a form similar to that studied by
K\"{o}nig {\it
et al.}
\cite{Konig} with no kinetic term:

\begin{eqnarray}
{\mathcal H} & = &
\ind{3}{r}\sum_{i,\alpha}J_{i\alpha}(\bvec{r})\bvec{s}_\alpha
\cdot
\bvec{S}_i(\bvec{r}) \nonumber \\
& & - \ind{3}{r} \left\{ g\mu_B \bvec{B}\cdot \sum_i
\bvec{S}_i(\bvec{r})
+ g^*\mu_B \bvec{B}\cdot \sum_\alpha\bvec{s}_\alpha(\bvec{r})
\right\},
\end{eqnarray}
where $\alpha$ labels the holes, $i$ labels the Mn spins,
$\bvec{s}_\alpha$ is the spin for
the $\alpha^{th}$ hole and $\bvec{S}_i(\bvec{r})$ is the Mn
spin centered on site $i$.  The
g-factors of the hole and Mn spins are labelled by $g^*$ and $g$
respectively, $\mu_B$ is the
Bohr magneton and $\bvec{B}$ is an external magnetic field,
which we shall assume to be zero unless otherwise stated.
The overlap integral for the $\alpha^{th}$ hole with the spin
centered on site $i$ is
$J_{i\alpha}$.  We assume that the d-electrons that give rise
to the Mn spins are localized in comparison to the holes and
treat the Mn spins as having
delta function spatial dependence, i.e. $\bvec{S}_i(\bvec{r}) =
\bvec{S}_i
\df{\bvec{r} - \bvec{R}_i}{3}$, which, after integrating out the
delta functions, leads to the Hamiltonian

\begin{equation}
{\mathcal H} = \sum_i \sum_\alpha
J_{i\alpha}(\bvec{R}_i)\bvec{s}_\alpha
\cdot \bvec{S}_i ,
\end{equation}
where $J_{i\alpha}(\bvec{R}_i) = J_0|\phi_\alpha(\bvec{R}_i)|^2$, with
$\phi_\alpha(\bvec{r})$ the
wavefunction of the $\alpha^{th}$ hole and $J_0$ the
microscopic exchange constant.
We can write $\sum_\alpha
J_{i\alpha}(\bvec{R}_i)\bvec{s}_\alpha =
\bvec{h}(\bvec{R}_i)$, and
treat the effect of the magnetization of the hole spins as
creating an
effective field at each Mn site, so our Hamiltonian is

\begin{equation}
\label{start}
{\mathcal H} = \sum_i \bvec{h}(\bvec{R}_i) \cdot \bvec{S}_i.
\end{equation}
In a recent mean-field study \cite{Mona2} of a tight-binding model of
the impurity band arising from holes on Mn sites coupled to
Mn spins, the randomness of the Mn sites is explicitly
taken into account. That work involves a numerical implementation of
the self-consistent mean field equations, in which the effective fields
are self-consistently calculated at {\em each} Mn site for each temperature, $T$.
(The mean-field in such treatments refers only  to temporal averaging over
the local environment, and not positional averaging).
A ferromagnetic phase is found below a critical temparture, with a
{\em distribution} of effective fields, $P(h)$, 
which is temperature dependent.  The numerical mean field model
explicitly includes the itinerant nature of the holes.  We use the
distribution of effective fields from the self-consistent mean-field calculations.
Hence, even
though the models constructed here are in the form of pure exchange,
the wavefunctions used to calculate effective couplings are those of
itinerant electrons. Consequently, that physics is implicit in these models.
In a mean field description
of carrier moments, the field is simply a product of the
mean moment (which we take to be along the $z$ direction, of
magnitude $s^z$) and an effective exchange interaction $J$ which varies
from site to site. Thus, we can obtain
from the numerical study, an effective distribution of
exchange couplings $P(J)$, and our Hamiltonian becomes an
integral over the distribution $P(J)$:

\begin{equation}
\label{dist}
{\mathcal H}_{\rm distribution} = \int_0^{J_{\rm max}} dJ \, P(J)
\, J \,
S^z_J s^z.
\end{equation}
Details of the derivation are given in Appendix \ref{derivation}.

In a fully self-consistent scheme (such as Ref. \onlinecite{Mona} )
the distribution $P(J)$ depends on the temperature $T$ due to the
temperature dependence of the hole amplitude distribution.
However, as we show below, the distribution $P_0 (J)$
calculated at $T = 0$ works very well for temperatures up to
$T = 0.6 \, T_c. $ Hence, in keeping with our search for a {\em simple}
model, we will use $P_0 (J)$, and consider the $T$-dependence
of $ P(J) $ only where necessary.

\subsubsection{Spin and carrier magnetization}
We find a self-consistent solution for the average Mn spin of

\begin{equation}
\left<S^z_J\right> = - SB_S(\beta\alpha_J),
\end{equation}
where $S = 5/2$ is the spin of each Mn ion,

\begin{equation}
B_S(x) = \frac{2S+1}{2S}\coth\left(\frac{2S+1}{2}x\right) -
\frac{1}{2S}\coth
\left(\frac{x}{2}\right),
\end{equation}
is the Brillouin function for spins with
magnitude $S$ and

\begin{equation}
\alpha_J =  J \left<s^z\right>.
\end{equation}
We also need to take into account the fact that the holes 
experience
an effective field from the Mn spins, which leads to the
following
self-consistency condition

\begin{equation}
\left<s^z\right> = -jB_j(\beta\alpha^*),
\end{equation}
where

\begin{equation}
\alpha^* = \frac{1}{p}
\int_0^{J_{\rm max}} dJ \, P(J) J \left<S^z_J\right>,
\end{equation}
$p$ is the ratio $n_h/n_{\rm Mn}$
and $j$ is the effective hole spin ($j = 3/2$ in real systems).
To
compare our results with the numerical study, \cite{Mona,Mona2}
we
use the Brillouin function  with $j = 1/2$ for the hole spin and
replace
$J_0$ by $3J_0$ so that in effect the hole spins take values of
$\pm 3/2$.
We emphasize however, that we have found that treating the spin
as a
classical object (i.e. using a Langevin function) or
using Brillouin functions with $j = 1/2$ or $j = 3/2$ does not
affect the basic picture we have here, or the characteristic
{\em shape} of the curves, other than in raising or lowering
$T_c$.
The value of $T_c$ is given by

\begin{equation}
\label{Tc}
T_c = \sqrt{\frac{35}{48} p \overline{J^2}},
\end{equation}
where $\overline{\cdots}$ denotes an average with respect to
$P(J)$
and $\left<\ldots\right>$ denotes a thermodynamic average.
Note that whilst $p$ enters Equation (\ref{Tc}) explicitly, there
is also an
implicit dependence, in that $p$ affects the distribution of
hole-Mn
exchange couplings $P(J)$ and hence $\overline{J^2}$.

\subsubsection{Susceptibility and Specific Heat}

We next calculate the magnetic susceptibility and specific heat
at zero field.  The susceptibility per unit volume is

\begin{equation}
\chi = \chi^* + \int_0^{J_{\rm max}} dJ\, \chi_J,
\end{equation}
where $\chi^*$ is the contribution to the susceptibility due to
the holes
and $\chi_J$ is the contribution to the susceptibility from Mn
spins
coupled
to holes with exchange $J$.  The individual
expressions for the susceptibilities are

\begin{eqnarray}
\label{hole_susc}
\chi^* & = & \frac{(g^*\mu_B)^2 n_h \beta G^*\left(1 -
\frac{g}{g^*}\frac{\beta}{p} \int_0^{J_{\rm max}}
dJ \, P(J) JG_J \right)}{1 - \frac{1}{p}\beta^2 G^*
\int_0^{J_{\rm max}}
dJ \,
P(J) J^2 G_J},  \\
\label{Mn_susc}
\chi_J & = & g\mu_B P(J) n_{\rm Mn} \beta G_J\left(g\mu_B -
\frac{1}{n_{h}}
\frac{J}{g^*\mu_B}\chi^*\right),
\end{eqnarray}
where

\begin{eqnarray}
\label{gdefa}
G^* & = & j\frac{d}{dx}\left[B_j(x)\right]_{x = \beta\alpha^*},
\\
\label{gdefb}
G_J & = & S\frac{d}{dx}\left[B_S(x)\right]_{x = \beta\alpha_J}.
\end{eqnarray}
We use the expressions for the susceptibility given in equations
(\ref{hole_susc}) and (\ref{Mn_susc}) with $g = 2$, and assume
$g^* = 2$
for our numerical calculations in section \ref{numerical}.
\cite{Schneider}

To obtain the specific heat we have to be 
careful since we
calculate with a Hamiltonian that has temperature
dependent coefficients
and the concentrations of strongly and weakly coupled spins may
also depend on temperature.  We use the mean field free energy
to obtain the specific heat through the relation

\begin{equation}
C_V = \frac{dU}{dT},
\end{equation}
where we note that the energy per Mn spin is

\begin{equation}
\frac{U}{N_{\rm Mn}} = \int_0^{J_{\rm max}}
dJ \, P(J) J \left<S_J^z\right>\left<s^z\right>,
\end{equation}
in our mean field theory, where $N_{\rm Mn}$ is the total number
of Mn
spins.

\subsection{Single and two component models}

For a system without disorder, the exchange between
carriers and spins can be characterized by a single coupling,
as has been done, {\em e.g.}, in Ref. \onlinecite{Konig}.
To demonstrate how these models are inadequate for the system
under consideration, we consider two simplified models
to substitute for the hierarchy of couplings implied by the
distribution $ P_0(J) $. The first has a single coupling
parameter, while the second incorporates the idea of ``strongly"
and ``weakly" coupled spins in terms of two coupling parameters.
In both cases, the parameters are determined from the distribution
$ P_0(J) $. We compare the results of these two models with
that obtained from the distribution $ P_0(J) $ in section \ref{numerical}
.

\subsubsection{Single component model}
In a single component model of Mn spins, where each spin
is coupled in the same way to the carrier spins,
the form for $P(J)$ is just a delta function:

\begin{equation}
P(J) = \delta(J - J_1),
\end{equation}
where $J_1$ is the exchange coupling.  In that
case, the
formulae for thermodynamic quantities derived for
a distribution in the previous section lead to

\begin{eqnarray}
\left<S^z\right> & = & - SB_S(\beta\alpha_1), \\
\left<s^z\right> & = & - jB_j(\beta\alpha^*),
\end{eqnarray}
with $\alpha^* = (1/p) J\left<S^z\right>$ and $\alpha_1 = J_1
\left<s^z\right>.$  The susceptibility per unit volume is

\begin{equation}
\chi = \chi^* + \chi_1,
\end{equation}
and

\begin{eqnarray}
\chi^* & = & \frac{(g^*\mu_B)^2 n_h \beta G^* \left( 1 -
\frac{g}{g^*}\frac{\beta}{p} J_1 G_1
\right)}{1 - \frac{1}{p}\beta^2 G^* J_1^2 G_1}, \\
\chi_1 & = & (g\mu_B) n_{\rm Mn} \beta G_1 \left( g\mu_B -
\frac{1}{n_{h}}
\frac{J_1}{g^*\mu_B}\chi^*\right),
\end{eqnarray}
where the $G$ functions have the same meaning as previously.
Finally we
calculate the specific heat again with the derivative of energy
with
respect to
temperature, but in this case the energy per Mn spin is

\begin{equation}
\frac{U}{N_{\rm Mn}} = J_1\left<S^z\right>\left<s^z\right>.
\end{equation}

\subsubsection{Two component model}
As we show in the next section, the calculated distribution
of exchange couplings are quite broad, and cover many orders
of magnitude for parameter values of interest. This motivates
us to study a model that is the next simplest after the single
coupling model, namely the two component model, with an
exchange distribution:

\begin{equation}
P(J) = \frac{n_1}{n_{\rm Mn}} \delta(J - J_1) + \frac{n_2}{n_{\rm
Mn}}
\delta(J - J_2).
\end{equation}
The physical motivation of the above distribution is to
divide the Mn spins into two types, one which is
strongly coupled to the carrier spins ($J_1$), with an effective
concentration $n_1$, and the other which is weakly coupled to
the carrier spins ($J_2$) with a concentration $n_2$. Since the
only energy scale characterizing the thermodynamics is the
temperature $T$, we would expect it to play an important
role in determining both the coupling constants $J_1$ and $J_2$,
and the concentrations of strongly and weakly coupled spins.
Consequently, we expect that the best fit to the curves for
a distribution of exchanges will be obtained when $J_i$ and $n_i$ 
are temperature dependent. This temperature dependence 
of parameters should
be viewed in the same spirit as in a variational fit to
free energies of actual (T-independent) Hamiltonians by
model Hamiltonians. \cite{feynman72}

We obtain a self consistent mean field solution to this model of

\begin{eqnarray}
\left< S^z_{a}\right> & =  & - SB_S(\beta\alpha_a), \\
\left<s^z\right> & = & - jB_j(\beta\alpha^*),
\end{eqnarray}
where

\begin{equation}
\alpha_a = J_a \left<s^z\right>,
\end{equation}
with $a = 1$ or $2$, and

\begin{equation}
\alpha^* = \frac{n_1}{n_{h}} J_1 \left<S^z_{i_1}\right> +
\frac{n_2}{n_{h}}J_2\left<S^z_{i_2}\right>.
\end{equation}

Using the notation introduced above, the susceptibility per unit
volume is

\begin{equation}
\chi = \chi^* + \chi_1 + \chi_2,
\end{equation}
where $\chi^*$ is the contribution to the susceptibility due to
the holes and $\chi_{1,2}$ are
the contributions from the two species respectively -- the
expressions for
the susceptibilities are

\begin{eqnarray}
\label{chihole}
\chi^* & = & \frac{(g^*\mu_B)^2 n_h \beta G^* \left(1 -
\frac{g}{g^*} \frac{n_1}{n_{h}}
\beta J_1  G_1 -
\frac{g}{g^*}\frac{n_2}{n_{h}} \beta J_2 G_2 \right)}{1 -
\frac{1}{p}
\beta^2 G^* (\frac{n_1}{n_{\rm Mn}} J_1^2 G_1 + \frac{n_2}{n_{\rm
Mn}}
J_2^2 G_2)},
\nonumber \\ \\
\label{chiMn}
\chi_a & = & g \mu_B n_a \beta G_a\left(g\mu_B -
\frac{1}{n_h}\frac{J_a}{g^*\mu_B}\chi^*\right),
\end{eqnarray}
where the $G$ functions are as defined in Equations (\ref{gdefa})
and
(\ref{gdefb}).
The energy per Mn spin (from which we determine the specific
heat) is

\begin{equation}
\frac{U}{N_{\rm Mn}} = \frac{n_1}{n_{\rm Mn}}J_1 \left<s^z\right>
\left<S^z_1\right>
 + \frac{n_2}{n_{\rm Mn}}J_2 \left<s^z\right> \left<S^z_2\right>.
\end{equation}

\begin{figure}[htb]
\centerline{\psfig{file=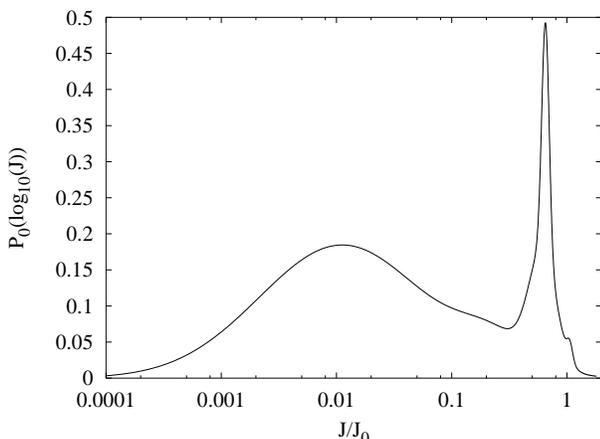,height=6cm,angle=270}}
\caption{The distribution $P_0(\log_{10} J)$ calculated for $p=0.1$ and
$x = 0.01$ in the numerical mean field model of Ref. 13
 at $T=0$.}
\label{PJ1}
\end{figure}

\section{Numerical tests of the one and two component models}
\label{numerical}
We now compare results obtained using our simple approximations
with the full numerical mean field calculations. We concentrate
on two cases, where the Mn concentration $n_{\rm Mn}$ leads to a
fractional
occupancy of the Ga sublattice by Mn ions of $x$ = 0.01 and 0.02.
For both cases, we consider a ratio
of hole concentration $n_h$ equal to 10\% of the Mn
concentration,
i.e. $n_h/n_{\rm Mn} = p = 0.1 .$  The distribution $P_0 (J)$
calculated
numerically \cite{Mona2} for the $x = 0.01$ case at $T = 0$,
using a Bohr radius  for the hole equal to 7.8
$\stackrel{\circ}{\rm A}$, \cite{Mona}
is shown in Fig. \ref{PJ1}.  As can be seen, it spans almost three orders
of magnitude, and
for this density, consists of two peaks. The higher peak is
due to sites where the exchange interaction is dominated by
a single hole that has a high probability of being on the Mn
site in question, whilst the lower peak is found to be due to
sites that have practically no amplitude for a hole, but whose
exchange field is coming from holes on nearby sites. As $x$ and
$p$ are changed, the relative weights in the two peaks changes,
as does the total width of the distribution. However, the
inferences made for this concentration remain valid for higher values
of $x$ (we have checked explicitly the cases $x = 0.02$ and $ x= 0.03 $).

\begin{figure}[htb]
\centerline{\psfig{file=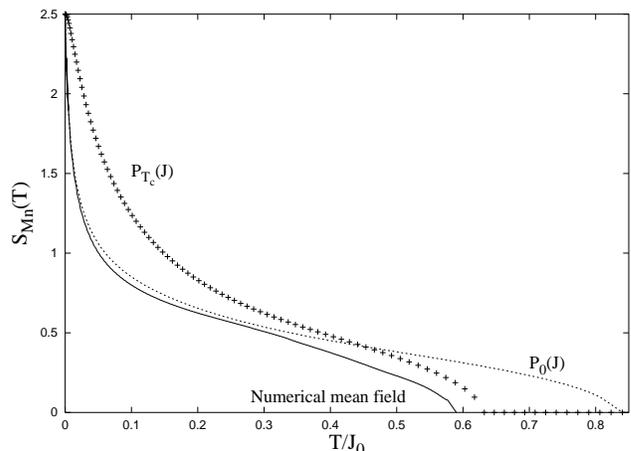,height=6cm,angle=270}}
\caption{Illustration of how the choice of $P(J)$  at different
temperatures for $x = 0.01$ influences the fit to the numerical mean field
magnetization
curve.}
\label{allthree}
\end{figure}

Figure \ref{allthree} plots the average Mn spin, $S_{\rm
Mn}(T)$, calculated using the self-consistent
solution of section \ref{Ham} with the distribution $P_0(J)$ (Fig.
\ref{PJ1}) 
(dashed line) and  the full numerical mean field result (solid line)
against temperature.
As can be seen clearly, the curve using the
$T=0$ distribution works very well (errors less than 1\%)
until about 60 \% of $T_c$ for the numerical mean field model.
To fit the numerical results properly
for higher $T$, one must allow for the
distribution of local fields to deviate from the
$T=0$ distribution as the polarization of
the holes begins to fluctuate strongly as the transition is
approached. In figure \ref{allthree}, we indicate with crosses
the $S_{\rm Mn} (T)$ curve obtained using the
distribution $P_{T_c} (J)$ (determined for $T=T_c$).  
Clearly, an interpolation of distributions at
the two extremes ($T=0$ and $T=T_c$) will be
adequate to reproduce the full numerical curve. \cite{footnote}

\begin{figure}[htb]
\centerline{\psfig{file=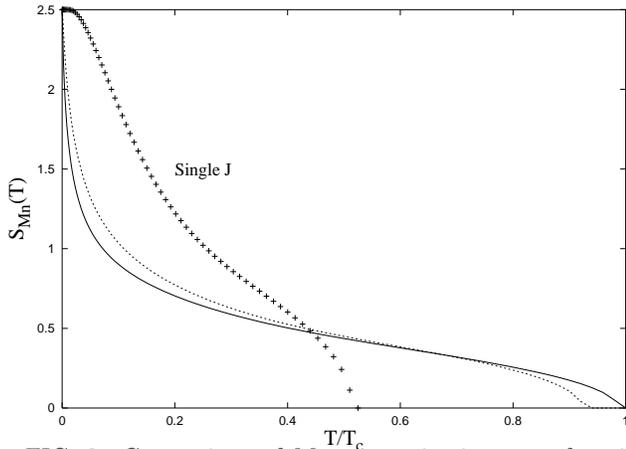,height=6cm,angle=270}}
\caption{Comparison of Mn magnetization as a function of temperature for
the
numerically derived
distribution $P(J)$ (solid line), the single coupling $J = \overline{J}$
(labelled,
crosses)
model and the two component model with linear averaging (dashed line)
where
$\gamma = 0.6$ and $x = 0.01$.}
\label{mag1}
\end{figure}

\begin{figure}[htb]
\centerline{\psfig{file=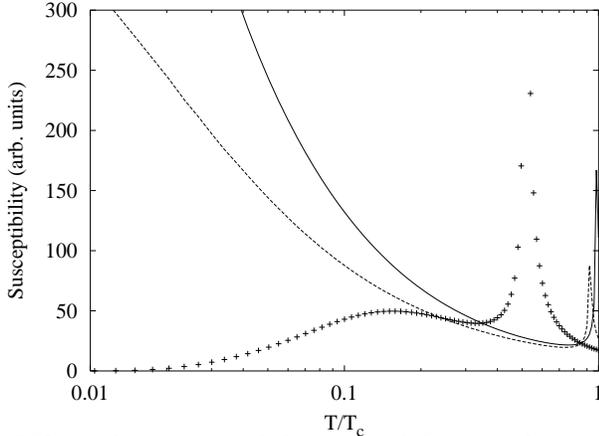,height=6cm,angle=270}}
\caption{Comparison of the susceptibility as a function of temperature
for $x = 0.01$ calculated
for the distribution $P(J)$ (solid line), the single coupling $J =
\overline{J}$ model (crosses) and the two component model (dashed line)
with $\gamma = 0.6$. Note that $T_c$ is that for $P_0(J)$.}
\label{chi}
\end{figure}

\begin{figure}[htb]
\centerline{\psfig{file=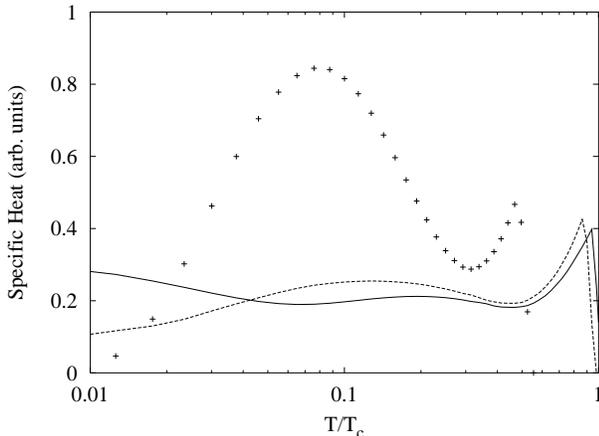,height=6cm,angle=270}}
\caption{Comparison of the specific heat as a function of temperature
for $x = 0.01$ calculated
for the distribution $P(J)$ (solid line), the single coupling $J =
\overline{J}$ model (crosses) and the two component model (dashed line)
with $\gamma = 0.6$. Note that $T_c$ is that for $P_0(J)$.}
\label{cv}
\end{figure}

\begin{figure}[htb]
\centerline{\psfig{file=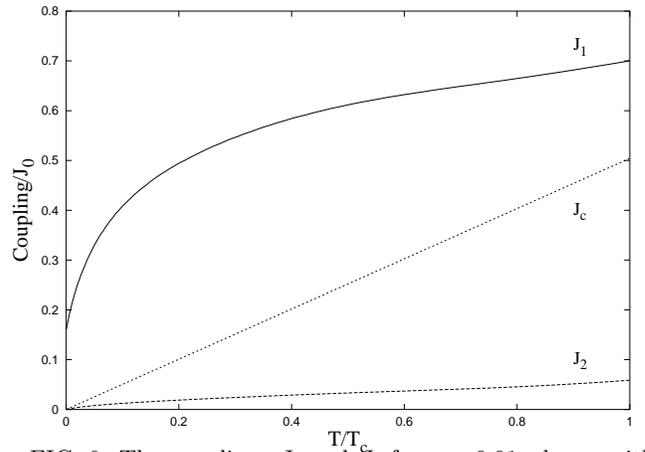,height=6cm,angle=270}}
\caption{The couplings $J_1$ and $J_2$ for $x = 0.01$,
shown with $J_c = 0.6 \, T$ as
functions of temperature where $J_1$ and $J_2$ are obtained in the
linear averaging scheme. Note that $T_c$ is that for $P_0(J)$.}
\label{coup}
\end{figure}

\begin{figure}[htb]
\centerline{\psfig{file=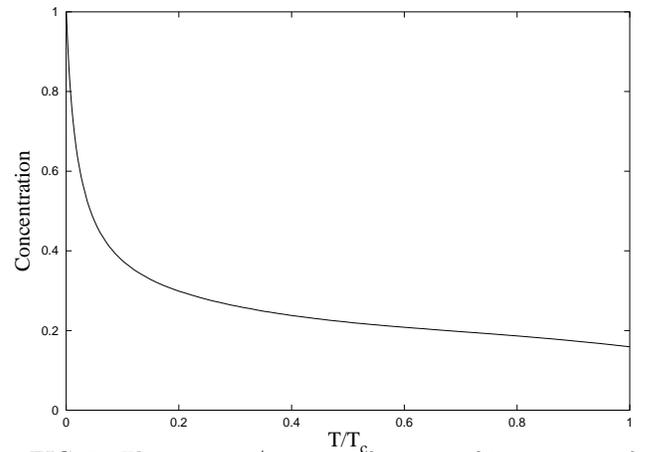,height=6cm,angle=270}}
\caption{The ratio $n_1/n_{\rm Mn}$ as a function of temperature for
$x=0.01$ using
the model in which $J_c = 0.6 \, T$ where $J_1$ and $J_2$ are
obtained in the
linear averaging scheme. Note that $T_c$ is that for $P_0(J)$.}
\label{conc}
\end{figure}

Our goal is, however, to simplify the description of
the distribution of exchange fields in
terms of a few couplings. To this end, we consider the
case with the fixed $T$-independent $P(J)$ shown in Fig.
\ref{PJ1},
and attempt to fit the thermodynamic properties
for that case using a one- or two-
component model of Mn spins. To fit the full
numerical results for temperatures up to $T_c$
(for the numerical mean field model), we would use
a similar scheme with the appropriate
$P(J)$ which best fits those numerical results.
In Figures \ref{mag1} to \ref{brillouin}, $T_c = 0.84 \, J_0$ refers to that
determined from $P_0(J)$ as shown in Figure \ref{allthree}. 

For the single coupling parameter model, we replace the
distribution $P(J)$ by the mean coupling

\begin{equation}
J_1 = \overline{J} = \int_0^{J_{\rm max}} dJ \, J P(J) ,
\end{equation}
which for $x=0.01$ when we use $P_0(J)$
gives $\overline{J} = 0.161 \, J_0$. \cite{footnote2}
Figure \ref{mag1} shows the average Mn spin  $S_{\rm Mn} (T)$
(dotted curves) calculated using this model for $x = 0.01$.
The results are clearly seen to be a
poor fit to the results obtained for the full distribution
(solid lines), over most of the temperature range.
\cite{footnote3}

In the two-component model, the parameters are determined
using the following scheme. First, since the temperature is
the only thermal energy scale in the problem that can be
used to define ``strongly'' and ``weakly'' coupled spins, we
define a cut-off coupling $J_c = \gamma T$ . 
All spins that have couplings below $J_c$ are weakly coupled,
whilst those with couplings greater than $J_c$ are strongly
coupled. 
Thus the concentrations $n_1$ and $n_2$ of the two sets
of spins are given by
the relative fractions of spins with couplings above and
below $J_c$, {\em i.e.}:

\begin{eqnarray}
\frac{n_1}{n_{\rm Mn}} & = & \int_{J_c}^{J_{\rm max}} dJ \, P(J),
\\
\frac{n_2}{n_{\rm Mn}} & = & 1 - \frac{n_1}{n_{\rm Mn}}.
\end{eqnarray}
On the other hand, the couplings $J_1$ and $J_2$ 
are taken to be the averages over
the two populations
\begin{eqnarray}
J_1 & = & \int_{J_c}^{J_{\rm max}} dJ \, JP(J) ,\\
J_2 & = & \int_{J_{\rm min}}^{J_c} dJ \, JP(J) .
\end{eqnarray}
In this scheme, the only adjustable parameter is the
dimensionless parameter $\gamma$, which determines $J_c$ and is chosen to
give the best overall fit to $S_{\rm Mn}(T)$. Using $P_0(J)$ yields
$\gamma = 0.6$. Since most thermodynamic functions depend
exponentially on the ratio $ J/T $, we expect this scheme
to work especially well when the distributions are broad,
and the demarcation between ``strongly'' and ``weakly'' 
coupled spins becomes sharp. \cite{Bhatt}
For narrow distributions, on the other hand, this scheme
essentially reduces to a single coupling, which should be
adequate for most purposes. 

The curve of Mn spin versus temperature $S_{\rm Mn} (T)$, obtained using
the
two-component
model, is shown in Figure \ref{mag1}
as a dashed line. In contrast to the
single coupling model the curve
has the same shape as for the full distribution
and provides a much better quantitative fit.
{\em It might appear
that using two couplings to approximate the distribution shown in
Fig. \ref{PJ1} works well because of the double peaked nature of
the
distribution.  We wish to emphasize that this is not the case;
the main reason actually appears to be the large width
of the distribution $P_0(J)$.  In particular, if the upper peak in $P_0(J)$
is
removed, then the magnetization
curve from the modified $P(J)$ still needs a
two-component model to provide an adequate fit and a one-coupling
parameter
fit works barely better than for the distribution shown in Fig. \ref{PJ1}.}

We now turn to other thermodynamic quantities,
the susceptibility and specific
heat; these quantities are shown for $x=0.01$ and were calculated 
for a distribution, as well as the
single and two component models,
 in Section \ref{mean_field}. 
Figure \ref{chi} plots the susceptibility for
the mean field model with the distribution $P(J)$, the two
component
model with $\gamma = 0.6$ and the single component model for $x =
0.01$.
Because the susceptibility is a higher order derivative,
the two component model is not quite as accurate as for the
magnetization.  Nevertheless, there is good agreement
with the results for the full distribution on a semiquantitative
level down to 10 \% of $T_c$ as obtained from $P_0(J)$,
whilst the single coupling
model bears little resemblance to the distribution.

Similarly, the curves for specific heat (Fig. \ref{cv})
as a function of temperature show good agreement
between the two-component model and the distribution
(in fact, better than for the
susceptibility) for temperatures greater than $ 0.1 \, T_c$.
In contrast, there is a strong quantitative discrepancy
between the single
coupling model and the distribution. The Schottky type
anomaly is broadened out considerably for the distribution
as well as the two-component model;
similar broadening is present in the full numerical
solution. \cite{Mona}

Figure \ref{chi}  apparently shows a
divergence in the susceptibility at low temperatures -- this
is not a divergence, but a peak at very low temperatures, due to
the considerable population of Mn spins with very small
local fields.

Finally, to understand better the
behaviour of the two-coupling model, we show the temperature
dependence
of the couplings $J_1$, $J_2$ and $J_c$ and also the temperature
dependence of the ratio $n_1/n_{\rm Mn}$ in Figures \ref{coup}
and
\ref{conc}
respectively.  

As expected from the physically motivated
criterion described earlier in this section,
at low temperatures nearly all
of the spins are strongly
coupled; however, as the temperature rises the number of strongly
coupled
spins decreases sharply, as the strength of the coupling
required for a spin to be strongly coupled increases with $T$.
It is this
essential aspect of a broad distribution of couplings that allows
the
two-component model to properly capture this behaviour.

\section{Discussion and Conclusions}

Whilst other authors have acknowledged that disorder plays some
role in the physics of
III-V DMS systems,
\cite{DOM,Schliemann,Boselli,SchliemannMC}
the results of mean-field \cite{Mona,Mona2} as well as Monte
Carlo simulations \cite{Xin1,Xin2} on models with a random
distribution of dopants 
near the metal-insulator transition
suggest that there should be strong effects
on the thermodynamic properties of DMS systems, especially at low
temperatures.
The tendency in experimental fits, on the other hand, has been
to neglect disorder effects entirely
and produce fits characteristic of homogeneous systems.
Thus, magnetization curves
obtained from transport
measurements in a sample with $x=0.035$, \cite{Ohno3}
are fitted with Brillouin functions to extract a single
exchange coupling. It is important to
ask whether the effects described here have been seen or have
the potential  to be seen
experimentally.  Significant
deviations from a Brillouin curve appear to be seen in
a number of experiments, \cite{Beschoten,Ohnoxx,Awschalom} in contrast
to Ref. \onlinecite{Ohno3}.  The Brillouin function-like
behaviour appears to arise in transport, whilst the deviations are
seen in SQUID measurements of the magnetization.  One explanation
for this may be that the transport measurement mainly samples Mn spins
that are strongly coupled to holes, which can give a Brillouin function as
described below.
However, the SQUID measurement
samples all Mn moments equally, and the magnetization inferred is
from both weakly and strongly coupled Mn spins (as is calculated here).
Interestingly, a measurement of the magnetization against
magnetic  field at a temperature of 2 K  ($T_c$ was 37 K)
in a sample with $x = 0.02$ showed that the magnetization was
roughly only 25 \% of
its saturation value. \cite{Ohldag} This is consistent with
our picture here that due to a broad
distribution of exchange couplings
there are significant
numbers of spins that are not polarized, even at low temperatures. 
In studies of insulating
samples it has been observed that the saturation moment was
consistent with only about 40\% to 50\% participation of Mn spins.
\cite{Oiwa,Nojiri}

\begin{figure}[htb]
\centerline{\psfig{file=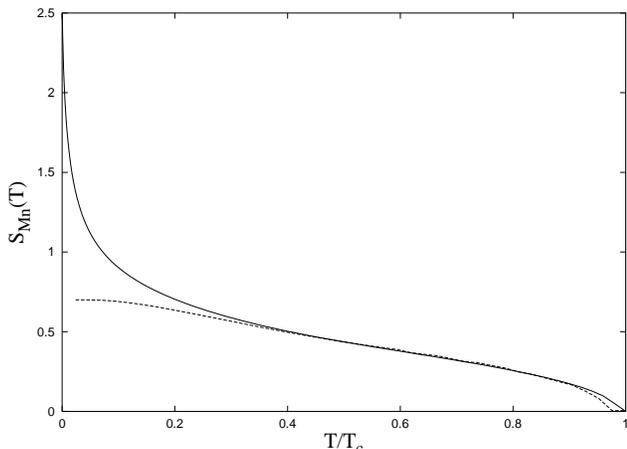,height=6cm,angle=270}}
\caption{Illustration of how a single Brillouin function can
approximate part of the magnetization curve of a distribution
($x=0.01$).  The parameters chosen for the
single Brillouin function were that 28 \% of the Mn spins were
coupled to holes with $J = 0.57 \, J_0$, whilst the rest were uncoupled.}
\label{brillouin}
\end{figure}

Figure \ref{brillouin} illustrates
how it is possible to have the magnetization from the
distribution $P_0(J)$ fitted
by a single Brillouin function over some temperature range.
The parameters used to
obtain the curve shown were to assume that only 28 \% of the Mn
spins contribute
to the magnetization and that the coupling is $J = 0.57 \, J_0$.
Also, in the large $x$ insulating
phase, there has been an observation
of a magnetization curve that could not be fitted with a
Brillouin function assuming only a
single coupling. \cite{VanEsch,VanBockstal}  In that work this
was ascribed to multiple exchange
mechanisms rather than disorder in the local fields as we suggest
here. We believe similar effects will persist
on the metallic side of the metal-insulator transition,
and the assumption of a single exchange coupling used in
experimental fits will be valid only deep in the metallic phase. 

\begin{figure}[htb]
\centerline{\psfig{file=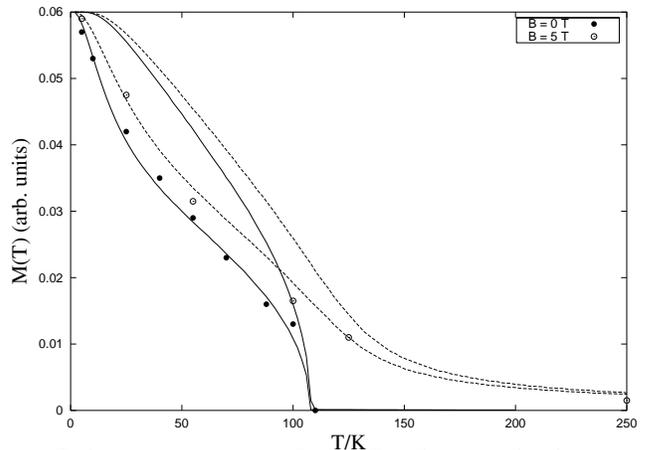,height=6cm,angle=270}}
\caption{Two component fit to SQUID magnetization curve at zero field and 
$B = 5$ T for a sample 
with $x = 0.053$ and $p=0.3$ from Ref. 27.  The
experimental data ($\bullet$ for $B = 0$ T, and $\circ$ for $B = 5$ T) is  fitted very well
with a two component model with $J_1 = 47.5$ K, $J_2 = 7.5$ K and $n_1 = 0.41 n_{Mn}$ (solid 
bold line for $B$ = 0 T, dashed bold line for $B$ = 5 T).  The light 
curves are for a single coupling and $J = 31$ K (solid $B$ = 0 T and dashed $B$ = 5 T).}
\label{ohnodata}
\end{figure}

To illustrate this point, we show  data for the 
remanent magnetization and field dependent magnetization
from a SQUID measurement (Ref. \onlinecite{Ohnoxx})  for
a sample with $x = 0.053$ and $p = 0.30$  that has been fit with $j=3/2$ at zero field 
and at $B = 5$ T.  (Note that the only change 
required to include a magnetic field in 
the formalism in Section \ref{mean_field}, is $\alpha \to \alpha - g\mu_B B$).
Since we obviously do not know $P(J)$ in this case, and we expect it to be
less broad in the metallic phase, we have taken temperature independent
$J_1$, $J_2$ and $n_1$, and compared the fit to one with a single $J$ that 
reproduces $T_c$ correctly.  Using $J_1 = 47.5$ K, $J_2 = 7.5$ K and $n_1 = 0.41 n_{Mn}$ 
we obtain the fits shown in Fig. \ref{ohnodata} as bold solid and dashed lines, which are in close accord
with the data at both magnetic field strengths, 
whereas a fit with a single $J = 31$ K (light and solid dashed lines) is clearly inadequate. 
Whilst a single fit of this form does not necessarily imply our model, it 
would clearly be of use to determine if a such a two component behaviour is universally seen,
by fitting experimental data for 
$M(H,T)$ at several fields and temperatures.  

In conclusion, we have found that we can approximate the
results of the numerical mean field treatment of a kinetic-
exchange model of Mn dopants in GaAs, which properly treats
the positional disorder in the alloy system, by a simple
exchange model with temperature-dependent effective couplings
between carriers and two different species of Mn spins.
Such a model emerges naturally because of the distribution
of couplings that are a consequence of the positional
disorder of the Mn ions.
Since the ultimate inputs in the simple model are
the couplings $J_i$ and concentrations $n_i$ of each
species, by parametrizing the behavior shown in Figures
7 and 8 in terms of a single parameter (or perhaps two),
experimental data can be used to yield information about
the true nature of coupling distributions in actual DMS
materials.
We hope that with more
detailed measurements, both of bulk magnetic and thermodynamic
quantities, as well as those from local probes, it may
be possible to parametrize the couplings accurately enough
to provide a quantitative fit to the other thermodynamic
properties at various concentrations in all regions of
the phase diagram, in the quantitative manner possible
for crystalline systems with translational symmetry.

\section{Acknowledgements}

We acknowledge many useful discussions with Xin Wan.
This project was supported by NSF DMR-9809483. MB
acknowledges Postdoctoral Fellowship support from
the Natural Sciences and Engineering Research Council
of Canada. RNB
thanks the Isaac Newton Institute, Cambridge University,
and the Aspen Center for Physics for hospitality
during the initial stages of this work.

\begin{appendix}
\section{Derivation of integral over effective couplings}
\label{derivation}
The Hamiltonian (Eq. (2)) may be rewritten as

\begin{equation}
{\mathcal H} = J_0 \sum_{i\alpha} |\phi_\alpha(\bvec{R}_i)|^2
\bvec{s}_\alpha
\cdot \bvec{S}_i,
\end{equation}
where we have used the definition $J_{i\alpha}(\bvec{R}_i) = J_0
|\phi_\alpha(\bvec{R}_i)|^2$.  We assume that
the
amplitude of the $\alpha^{\rm th}$ hole at site $i$ may be written as the
sum
over Mn sites of the product of the amplitude for a hole to be at a given site
times the amplitude from the local atomic orbital $\psi(r)$ (note that 
$\psi(r) \propto e^{-r/a_B}$).  Hence

\begin{equation}
|\phi_\alpha(\bvec{R}_i)|^2 = \sum_j p_\alpha(\bvec{R}_j)|\psi(\bvec{R}_i
- \bvec{R}_j)|^2,
\end{equation}
where we are in effect ignoring any quantum interference terms.  This
allows us
to rewrite the Hamiltonian as

\begin{equation}
{\mathcal H} = J_0 \sum_{ij\alpha} p_\alpha(\bvec{R}_j) |\psi(\bvec{R}_i
- \bvec{R}_j)|^2 \bvec{s}_\alpha
\cdot \bvec{S}_i.
\end{equation}
One of our mean field assumptions (based on the idea that each hole
interacts
with many Mn spins \cite{Mona2}), is that each hole behaves identically --
in
which case, $\left<\bvec{s}_\alpha\right> = \left<s\right>$ independent of
$\alpha$.
This also means that $p_\alpha(\bvec{R}_j) = (1/N_h) p(\bvec{R}_j)$,
where
$p(\bvec{R}_j)$ is the amplitude for finding {\it any} hole at site $i$.

If we define

\begin{equation}
J_i = J_0 \sum_j p(\bvec{R}_j)|\psi(\bvec{R}_i - \bvec{R}_j)|^2,
\end{equation}
then the effective exchange field at site $i$ is $J_i$ and the Hamiltonian
takes the form
\begin{equation}
{\mathcal H} = \sum_{i} J_i \bvec{s}\cdot\bvec{S}_i.
\end{equation}
To convert this sum into an integral, define

\begin{equation}
P(J) = \frac{1}{N_{\rm Mn}}\sum_i \delta(J - J_i).
\end{equation}
In mean field, $\left<\bvec{S}_i\right>$ depends only on $J_i$, hence 
we
can relabel $S_i = S_J$ and the Hamiltonian takes the form of Eq. (4) when
we
use the identity

\begin{equation}
\int_0^{J_{\rm max}} dJ \, P(J) = 1.
\end{equation}

\end{appendix}


\begin{references}
\bibitem{footnote} Interestingly, because spin 3/2 holes
lead to a lower $T_c$ than spin 1/2 holes,
the magnetization curves for spin 3/2 holes
using the distribution $P_0 (J)$
overlap almost perfectly with the numerical mean field theory
(with spin-1/2 holes) for all $T$.
\bibitem{footnote2} It may appear from the equation for $T_c$,
(Eq. (\ref{Tc})) that we should fit the r.m.s exchange coupling.
This would lead to minor modifications to $T_c$; however the significant
difference in shape of $S_{\rm Mn}(T)$ curves cannot be rectified
by adjusting $T_c$.
\bibitem{footnote3} The same quality of fit is seen for $x = 0.02$
or $x = 0.03$.
\end{references}
\end{document}